\documentclass[12pt]{article}
\usepackage[utf8]{inputenc}
\usepackage{xcolor}
\usepackage{float}
\usepackage[justification=centering]{caption}
\usepackage[margin=0.5in]{geometry}
\usepackage[pdftex]{graphicx}
\graphicspath{{\Documents}}     
\DeclareGraphicsExtensions{.pdf,.jpeg,.png}
\usepackage{titlesec}
\usepackage{moreverb}
\usepackage{makecell}
\usepackage[utf8]{inputenc}
\usepackage{multicol}
\usepackage{amsmath,amssymb,amsfonts}
\usepackage{algorithmic}
\usepackage{textcomp}
\usepackage{cite}
\usepackage{multirow}

\title{\textbf{Optimal Load Shedding for Public Safety Power Shutoffs}}

\begin{multicols}{2}
\author{Aniruddha Rajendra Rao\\
Industrial AI Lab\\
Hitachi America, Ltd., R\&D\\
Santa Clara, USA\\
Aniruddha.Rao@hal.hitachi.com\\
\and
Chandrasekar Venkatraman\\
Industrial AI Lab\\
Hitachi America, Ltd., R\&D\\
Santa Clara, USA\\
\vspace{1cm}
Chandrasekar.Venkatraman@hal.hitachi.com\\
\and
Robert Ellis\\
CDS-DSL\\
Hitachi Limited\\
Kokubunji, Tokyo, Japan\\
\vspace{1cm}
robert.ellis.ab@hitachi.com\\
\and
Chetan Gupta\\
Industrial AI Lab\\
Hitachi America, Ltd., R\&D\\
Santa Clara, USA\\
\vspace{1cm}
Chetan.Gupta@hal.hitachi.com\\}
\end{multicols}

\date{}

\begin{document}

\maketitle
\vspace{-1.5cm}
\hspace{0.35cm}\textbf{Keywords:}   Optimization, Genetic Algorithm, Load Distribution, PSPS, Load Shedding.

\abstract{Public utilities are faced with situations where high winds can bring trees and debris into contact with energized power lines and other equipments, which could ignite wildfires. As a result, they need to turn off power during severe weather to help prevent wildfires.  This is called Public Safety Power Shutoff (PSPS). We present a method for load reduction using a multi-step genetic algorithm for Public Safety Power Shutoff events. The proposed method optimizes load shedding using partial load shedding based on load importance (critical loads like hospitals, fire stations, etc). The multi-step genetic algorithm optimizes load shedding while minimizing the impact on important loads and preserving grid stability. The effectiveness of the method is demonstrated through network examples. The results show that the proposed method achieves minimal load shedding while maintaining the critical loads at acceptable levels. This approach will help utilities to effectively manage PSPS events and reduce the risk of wildfires caused by the power lines.}

\section{Introduction}

Strong winds can cause tree branches and debris to come in contact with active power lines, resulting in equipment damage and an increase in the risk for wildfires. Hence, utility companies use Public Safety Power Shutoff (PSPS) during severe weather to prevent wildfires. The current methodology involves forecasting of severe weather up to a week before power is shutoff and notifying customers two days in advance. Efficiently determining areas for power shutoff is challenging. Early shutoff causes revenue loss, while delaying increases wildfire risk. Recent tragic wildfires highlight the need for improved strategies due to electrical faults causing significant damage. For example, in the year 2011, CBC \cite{CBC} reported two wildfires in Bastrop county that got started by trees coming in contact with nearby power lines and that became the most destructive wildfires in Texas's history. It resulted in four deaths and caused over \$300 million in damage. In California, the 2018 Camp Fire, which was ignited by a power line, led to the death of 84 people and caused approximately \$9.3 billion in residential property damage alone \cite{r1}. The severity of this and other fires that occurred during the 2017 and 2018 California fire seasons led Pacific Gas \& Electricity (PG\&E), the responsible utility company, to file for bankruptcy and face charges for involuntary manslaughter \cite{pg1, pg2}. In 2021, it was confirmed that California’s Dixie Fire – the second-largest wildfire in California’s history - was sparked when power lines owned by Pacific Gas and Electric (PG\&E) came into contact with a tree. The Dixie fire went on to burn over 950 thousand acres across five counties and lasted almost 4 months. The more recent wildfires on going in different parts of North America and Europe are causing catatrophic losses. There is a need to effectively mitigate such situations.

The likelihood of wildfires caused by power infrastructure is higher during windy conditions, which can cause power failures and make wildfires more difficult to contain. Studies show that wildfires ignited by power lines are usually larger and more damaging than others \cite{MILLER2017267}. Power infrastructure can cause ignitions in various ways, with the most common being contact between vegetation and conductors \cite{f4}. Although efforts have been made to reduce the probability of ignitions, day-to-day operations leave utilities with limited options to reduce the wildfire risk. The only way to completely avoid the risk of ignitions is to de-energize the power line, this intentional black outs are referred to as Public Safety Power Shutoffs (PSPS) \cite{pg3, cec}. But this impacts the power system's ability to provide reliable electricity and can result in intentional blackouts affecting millions of people \cite{f5}. Our goal is to mitigate such a situation with optimal load shedding.

Load shedding is a crucial aspect of power system management during extreme conditions, such as wildfires, storms or any other calamity, where there is a risk of power failures and potential fire ignitions. Over the years, researchers have focused on optimizing different approaches for load shedding and energy generation to reduce the impact of power failures \cite{o2, o3, o4}. However, there is still a gap in addressing the problem of load shedding for public safety power shut-offs (PSPS) with partial load shedding while also considering load importance. Despite the availability of various optimization approaches, none of them tackles these specific challenges associated with PSPS, which can have significant economic and societal impacts. Hence, it is crucial to address these challenges to ensure that load shedding is conducted in a way that minimizes the adverse impact on the public while still maintaining safety during extreme conditions.


Our study addresses utility companies' challenges in managing PSPS while minimizing load shedding. It lays the basis for a tool aiding incident managers to identify optimal load shedding while maintaining grid stability. Our novel contributions include an optimization framework for minimizing wildfire and cascading failure risks during transmission line failures, partial load shedding considering load importance, and demonstrating the approach's efficacy using the RTS-GMLC test case under N-1 contingency. The paper's structure comprises problem formulation and our load shedding approach details in Section 2, RTS-GMLC case study in Section 3, and concluding remarks with future research directions in Section 4.

\section{Optimal Load shedding}

One of the most accurate way to solve the load shedding problem would be through discrete/binary combinatorial optimization. In this, we generate all possible solutions and then select the best one according to a predefined fitness function (objective function). For problems with "N" variables, the total number of solutions that need to be generated is $2^{N}$ for binary optimization. However, generating so many solutions takes time, and the limitations of computational devices further restricts solutions generation \cite{b1}. Traditional deterministic approaches are rigid and don't factor uncertainty. To overcome this, recent studies have proposed stochastic methods such as simulated annealing, ant colony, particle swarm optimization (PSO), and genetic algorithm (GA) to reduce the time required to find the best solution. Among these methods, GA \cite{o1} is effective in solving discrete as well as binary combinatorial optimization problems. The GA is capable of real-time applications as it converges rapidly.

\begin{figure}[thb]
    \centering
	\includegraphics[width=0.7\linewidth]{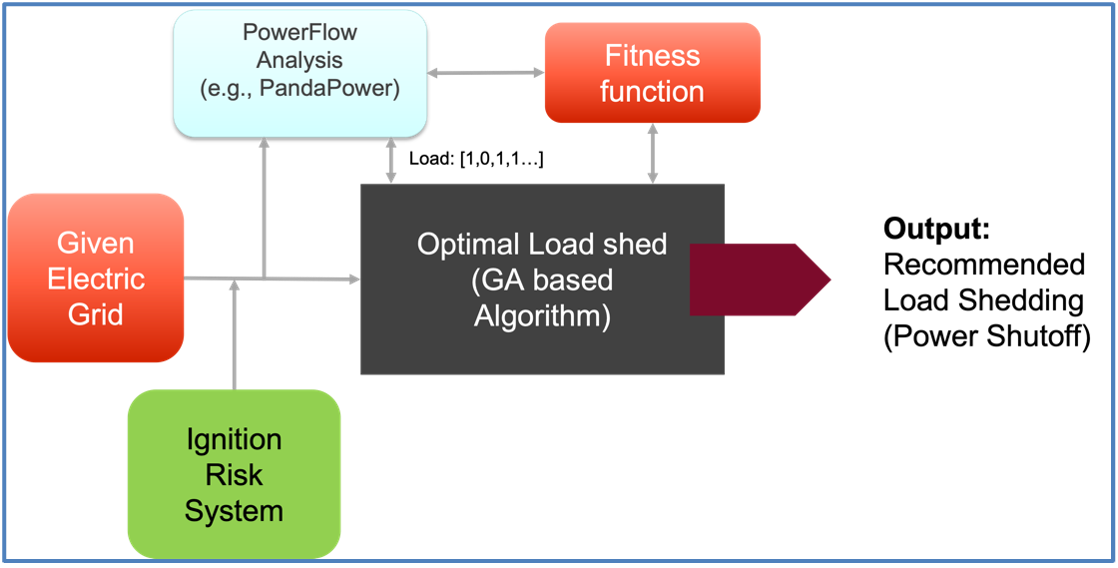}
	\caption{Overview of our Solution}
	\label{fig1}       
\end{figure}

Therefore, our research is based on applying GA for optimal load shedding for PSPS and validating it with grid networks, primarily for transmission lines \cite{ir1, ir2}. We further enhanced the research output to address partial load shed, i.e. instead of completely turning off a load in a bus, to shed partial load, thus achieving minimal load shed.  The second enhancement is related to ‘load importance’ – while minimal shedding is the desired state, it is also important to consider where the power is being consumed – such as hospitals, critical infrastructure, call centers, fire stations, etc. Optimal load shedding given load importance is crucial to minimise the adverse economic and societal impact of the load shedding on consumers \cite{i2, JABIAN2018486}. Our research addresses how to provide recommendations to incident commanders once they have an idea of ignition risk to optimally choose which loads to shed, as shown in Figure \ref{fig1}.

In Genetic Algorithm (GA) we have to define a fitness function 
that we want to maximize. We would like to define a fitness function such that the system is stable and load shedding is minimal and practical. The algorithm execution involves several steps.

In Step 1, a sample of the load system is taken from the grid network using a power flow analysis tool to act as the setup for the GA. The GA first initializes a population with N random solutions for optimization. Each bus with load (called as gene) in the network is initialized with a random load value from unit range. These random solutions are called chromosomes. The length of the chromosome is equal to the number of loads in the system.

In Step 2, the fitness function is evaluated for each chromosome and ranked accordingly. The most fit chromosomes (parents) are selected from the population to generate the offspring for the next generation. Number of parents selected will be M. Typically, this is done using some selection strategy, such as tournament selection or roulette wheel selection.


In Step 3, offspring solutions (chromosomes) are created through crossover operations, such as single-point or multi-point crossover. In single-point crossover, genetic material between two parents is swapped at a random point, yielding two offspring solutions with mixed genetic material. In multi-point crossover, multiple points are chosen for swapping, resulting in even more diverse offspring solutions compared to single-point crossover.

In Step 4 we perform mutation. A small random change is introduced into the offspring solutions to explore new regions of the search space. This is typically done using a mutation operator, which randomly alters a small portion of the offspring's genetic material.

In Step 5, the fitness function is computed for the new chromosomes. Only the top N fittest chromosomes remain in the population as the new generation. Step 3 and Step 4 is repeated till specified number of iterations are completed or stopping criteria is met. The chromosome with the lowest load shedding in the final generation is selected as the best solution. 

For our approach, we set the values of most genes in each chromosome as 1, since ideally we would like to have all the loads operating at full capacity. This helps us to reach the solution faster. We use single-point crossover and a mutation factor of 0.1. A load with value 0 means that is it completely shut, a load with value 1 implies it is at 100\% capacity and a load with 0.8 value means it is operating at 80\% capacity.

\begin{figure*}[h]
	\centering
	\includegraphics[width=170mm]{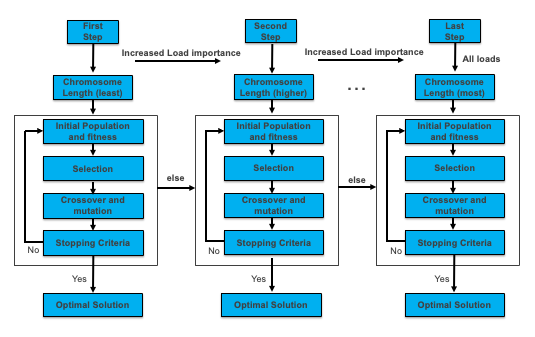} 
	\caption{Our multi-step partial load genetic algorithm approach for optimal load shedding}\label{Fig:arch}
        \label{Fig10}
\end{figure*}

We incorporate load importance using multi-step genetic algorithm approach. Multi-step GA is a type of optimization algorithm that uses a combination of GA and multiple stages of optimization to solve complex problems. In the context of optimal load shedding, multi-step GA can be used to optimize load shedding for multiple loads with different priorities (importance) as seen in Figure \ref{Fig10}. We generate load importance values randomly using a standard distribution but in practice, the incident managers will use domain knowledge and experience to specify load importance.

The first step of the multi-step GA is to utilize the load importance values to define different stages or levels of optimization, with each stage focusing on a specific subset of loads based on their importance.

For example, in a power network with multiple loads, some loads may be more critical than others in terms of their importance to the overall functioning of the grid. By assigning importance values to each load, a multi-step approach can be designed to optimize the loads in stages, with the least important loads being optimized for shedding first, followed by the next set of loads with increasing importance if necessary.

In the first stage, the least important loads are optimized using a GA, while the other important loads are assumed to be 1 (functioning at full capacity), with the fitness function being defined to make sure the network is stable and has maximum load capacity (or minimum load shedding). Once the optimal solution for this stage is found, the loads are adjusted accordingly in the network to make it stable otherwise we move on to the next stage.

In the second stage, the next set of important loads are optimized along with the least important loads, meaning the number of genes (loads) in a chromosome has increased slightly, while keeping the more important loads set to 1. This process is repeated for each subsequent stage until a solution is reached or till the last stage where all the loads are included to find a solution. Each stage is optimizing a higher subset of loads based on their importance.

By breaking down the optimization process into stages based on load importance, the GA can be run sequentially to optimize the loads in a more efficient manner. This approach also ensures that the most critical loads are given priority in the optimization process, leading to a more reliable and stable power grid. The main advantage of a multi-step GA is that it can handle complex optimization problems that are difficult to solve using traditional optimization methods. However, the algorithm requires careful design and tuning to ensure that it converges to a good solution in a reasonable amount of time.

\section{Case Study}

We assess our approach by conducting tests using a power system located in an area that faces a high risk of wildfires. We consider the RTS-GMLC \cite{f6} (Reliability Test System - Grid Modernization Laboratory Consortium) network, which is a widely used benchmark power system test case for research and development in the field of power system analysis and optimization. The test case was developed by the Grid Modernization Laboratory Consortium of the US Department of Energy, and it includes detailed modeling of the power grid, including generators, transmission lines, and loads.

\begin{figure}[thb]
    \centering
	\includegraphics[width=0.45\linewidth]{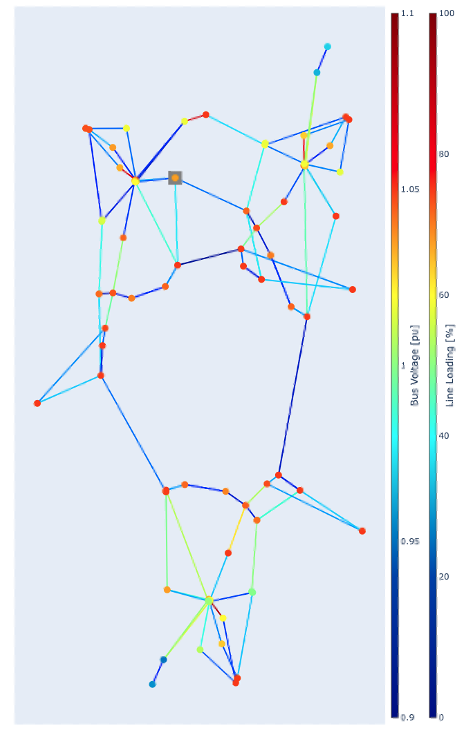}
	\caption{RTS-GMLC network with no power shutoff applied}
	\label{fig3}       
\end{figure}

The RTS-GMLC test case is based on the IEEE Reliability Test System (RTS) and includes additional features such as renewable generation, demand response, and energy storage systems. The RTS-GMLC test case includes 73 buses, 104 transmission lines (with the exception of the HVDC line being removed), 51 loads, 3 shunts and 18 generators as seen in Figure \ref{fig3}. The RTS-GMLC system is geographically located in southern California, Nevada, and western Arizona. These areas cover a lot of high risk wildfire areas.

The fitness function was designed to optimize for system safety and the remaining load in the system, with the maximum fitness value being 18726. It is defined as $Fitness Function=I*(10000)+L$, where $I$ is an indicator function for system safety and the $L$ is measuring the sum of the remaining loads (addition of Active and Reactive power at each load). A system is considered safe when none of the lines have a load percent more than 100 and the load bus voltage is between 0.95 to 1.05 of the nominal voltage. The safety criterion can be modified according to the problem in hand. We check for system safety using PandaPower \cite{pp}. It is a power systems analysis toolbox which is used for automation of analysis and optimization in power systems. The results obtained from this analysis can provide insights into the optimal load shedding strategies for maintaining system safety during line outages that can happen due to multiple reasons like wildfires, repairs, etc. We loop through the cases where each of the 104 lines in the network are down one at a time.

Table 1 shows the result of optimal load shedding under different approaches (Binary optimal load shedding and Partial optimal load shedding using GA) and conditions (different saturation values). We can use the saturate functionalities to stop the GA when the fitness value does not change for a defined number of iterations. We ran the GA for 500 iterations under no conditions and observed that for both approaches we are able to find solutions in 17 cases when a line is down, there is no feasible solution in 2 cases when a line is down, and for 85 cases there is no instability in the network when a line goes down. No feasible solution might be because of the limitation in pandapower to run contingency analysis for these 2 cases. This could be overcome by using Linear Sensitivity analysis \cite{jalpa} but this is not considered in this work. Our results are insightful, because mitigation measures, like adding tie lines, can be built for the 2 cases (Line 9 and Line 10) when there is a potential for cascading failure. This demonstrates how our approach can be used to determine the robustness of a power grid network. We get similar level of performance with saturation value of 25 slight increase in number of no feasible solution. The saturation value is a important hyper parameter which helps in increasing the speed but also decreases efficacy if the value is too low. From the last column in Table 1, we see that the total time to execute our approach on the 104 cases is minimal compared to using brute force. The performance of Binary vs Partial is similar though partial GA takes a bit longer to find the solution as we have increased the search space for each load from 0 or 1 to 0, 0.1, 0.2,….,1 (a step increase of 10\%). We can reduce the step increment even further to decrease the amount of load shed. Also, we our approach is more reliable than the incident manager, who takes actions based on just experience without any mathematical backing.

The fitness value is also higher for partial approach since we are considering partial load shedding. This is visible in Table 2, where we compare the load shed for each case when a line goes down and we must find an optimal load shedding solution. We can see from the table that the partial load shedding has a lower load shed amount (active power, p\_mw) every time compared to binary load shedding. In some cases, the gain in performance is huge.
This means that we can avoid unnecessary load shedding with this mechanism. The solution of partial optimal load shedding can be improved upon by running for large iterations over a broader search space.

\begin{table*}[]
\footnotesize
\centering
\caption{Results using Genetic Algorithm for N-1 contingency}
\begin{tabular}{|l||l|l|l|l|}
\hline
\makecell{Approach and Condition} &
  \makecell{Number of lines\\ for which there is\\ no feasible solution} &
  \makecell{Number of lines\\ for which there\\ is solution} &
  \makecell{Number of lines\\ for which there\\ is no instability} &
  \makecell{Total Run\\Time (seconds)} \\ \hline
Binary and no condition   & 2 & 17 & 85 & 1209 \\ \hline
Binary with Saturate=25   & 6 & 13 & 85 & 95   \\ \hline
Partial and no condition & 2 & 17 & 85 & 1763 \\ \hline
Partial with Saturate=25  & 3 & 16 & 85 & 17   \\ \hline
\end{tabular}
\centering
\label{tab1}
\end{table*}

\begin{table*}[htbp]
\footnotesize
\centering
\caption{Comparison of Partial load shedding vs Binary load shedding using Genetic Algorithm.}
\begin{tabular}{|l||l|l||l|l|}
\hline
  \makecell{Line down\\ (N-1 \\contingency)}  &
  \makecell{Total partial\\ load shed \\(MW)}  &
  \makecell{Load number associated\\ with partial load shedding\\ (load percent)} &
  \makecell{Total binary\\ load shed \\(MW)}  &
  \makecell{Load number \\associated with \\binary load shedding\\} \\ \hline
4  & 13.6 & 5  (0.9)  & 136 & 5       \\ \hline
8  & 40.8 & 5  (0.7)  & 136 & 5       \\ \hline
16 & 13.6 & 5  (0.9)  & 74  & 3       \\ \hline
18 & 34.2 & 7  (0.8)  & 171 & 7       \\ \hline
19 & 26.6 & 4 (0.9), 9 (0.9)  & 97  & 1       \\ \hline
23 & 54   & 2  (0.7)  & 97  & 1       \\ \hline
40 & 13.6 & 22 (0.9)  & 272 & 5, 22   \\ \hline
44 & 40.8 & 22 (0.7)  & 244 & 0, 22   \\ \hline
46 & 34.2  & 24 (0.8) & 261 & 5, 23   \\ \hline
47 & 34.2 & 24 (0.8)  & 331 & 9, 23   \\ \hline
73 & 13.6 & 39 (0.9)  & 136 & 39      \\ \hline
75 & 32.4  & 34 (0.7)  & 71  & 38      \\ \hline
76 & 42.6 & 38 (0.6)  & 74  & 37      \\ \hline
77 & 40.8 & 39 (0.7)  & 136 & 39      \\ \hline
79 & 25   & 40 (0.8)  & 171 & 41      \\ \hline
80 & 34.2 & 41 (0.8)  & 125 & 40      \\ \hline
90 & 72   & 36 (0.6)  & 180 & 36      \\ \hline
\end{tabular}
\centering
\label{tab2}
\end{table*}

\begin{table*}[htbp]
\footnotesize
\centering
\caption{Comparison of Partial load shedding vs Partial load shedding with importance using Genetic Algorithm.}
\begin{tabular}{|l||l|l||l|l|}
\hline
  \makecell{Line down\\ (N-1 \\contingency)} &
  \makecell{Total partial\\ load shed \\(MW)} &
  \makecell{Load number associated\\ with partial load shedding \\(load percent, importance)} &
  \makecell{Total parital \\and important  load\\ shed (MW)} &
  \makecell{Load number associated\\ with partial and important\\ load shedding\\ (load percent, importance)} \\ \hline
19 & 26.6 & 4 (0.9, 0.825), 9 (0.9, 0.959)  & 27.2 & 5 (0.8, 0.711)  \\ \hline
23 & 54   & 2 (0.7, 0.915)  & 75.6 & 0 (0.3, 0.794)  \\ \hline
80 & 34.2 & 41 (0.8, 0.900) & 37.5 & 40 (0.7, 0.734) \\ \hline
90 & 72   & 36 (0.6, 0.750) & 97.2 & 34 (0.1, 0.566) \\ \hline
\end{tabular}
\label{tab3}
\end{table*}

\begin{figure*}[h]
	\centering
	\includegraphics[width=170mm]{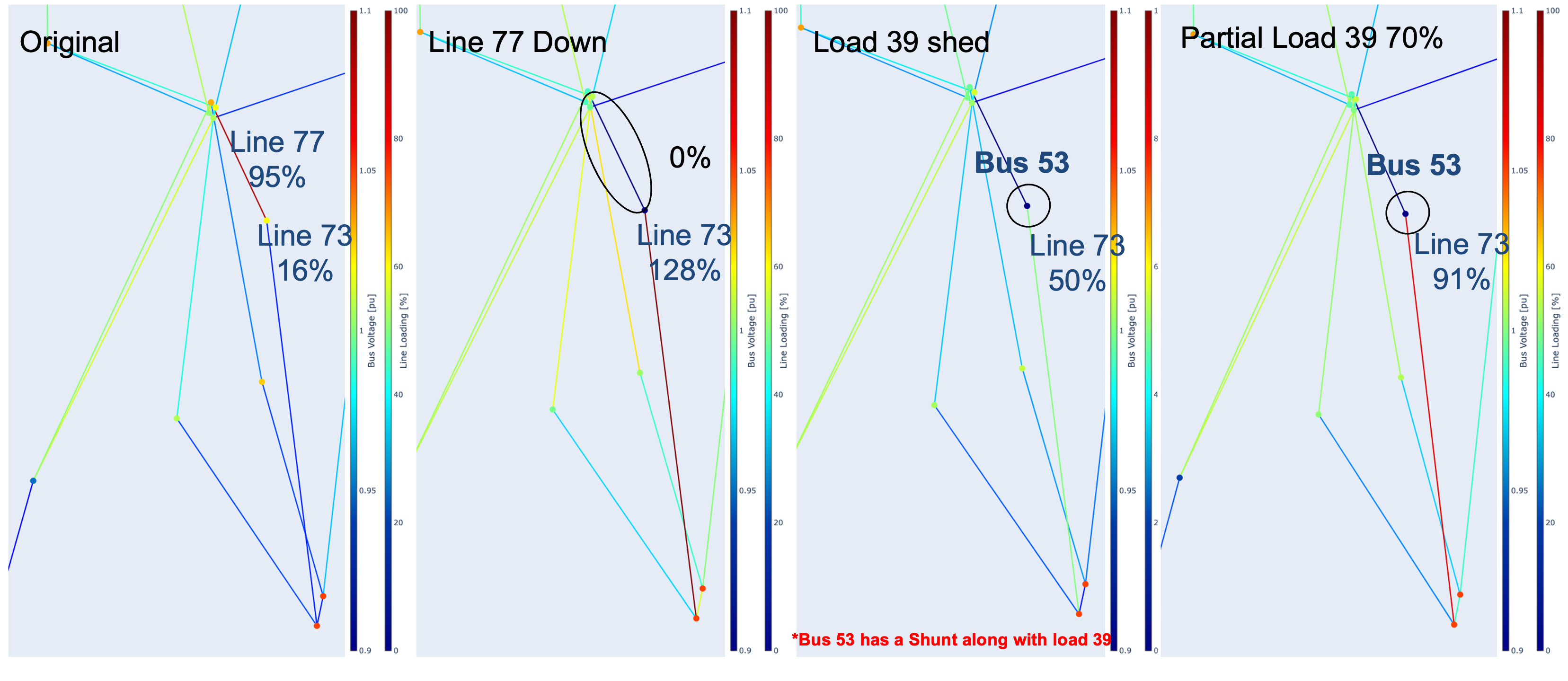} 
	\caption{Solution comparison when line 77 is down}\label{Fig:arch}
        \label{Fig4}
\end{figure*}

Figure \ref{Fig4} shows the network information when line 77 is down. We see the original network information on the left, followed by the network information when line 77 is down. It can be seen that line 73 is overloaded (line load is 128\%). The third sub-figure from the left shows the binary load shedding solution and the last one is for the partial load shedding solution. In both cases, load 39 at bus 53 is shut but in the partial case, load 39 is operating at 70\% capacity (in other words, 30\% load is shed) for the system to be stable, which is more optimal than completely shutting that load. Hence, partial load shedding helps in minimizing the economic and societal impacts while giving better service to the customers.

Next, we look into load importance. We assign random load importance to the 51 loads using beta distribution ($\alpha$=5 and $\beta$=1). In practice, these information will be know by the incident managers. In our Multi-step GA approach, in the first step, we only consider non-critical loads (load importance $<$ 0.8) for shedding. If we don't find an optimal load shedding solution in the first step, then we consider all the loads for load shedding in the second step. The performance efficacy is similar to Table 1 with slight increase in execution time. Table 3 shows how partial optimal load shedding performs with and without load importance. For example, we can see that when line 23 is down, and if we don't consider load importance, then the solution is to set load 2 at 70\% capacity resulting in 54 MW of load shedding but the load importance of load 2 is 0.915 whereas when we consider load importance, the solution is to set load 0 at 30\% capacity resulting in 75.6 MW load shedding and load 0 has lower importance (0.794) compared to load 2. Table 3 indicates that we are always successful in making sure no important load is shed if an alternate solution exists. Also, we are able to achieve this with marginal increase in load shedding. In the other cases, the load number shed is the same with or without importance consideration and hence not shown in the table. Also, we don't consider the binary case for load importance as the partial load shedding is superior. 


\begin{figure}[]
    \centering
	\includegraphics[width=0.6\linewidth]{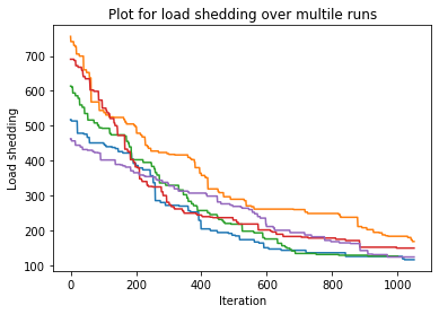}
	\caption{Load convergence of GA.}
	\label{fig2}       
\end{figure}

The convergence of the load shedding is demonstrated in Figure \ref{fig2}. The GA runs until the specified number of iterations is reached or a stopping criterion is met to make the network safe and the total load shed is minimized. We can see that the gains in load shedding after some iterations lead to minimal increase in performance and independent runs (denoted using different colors) give a similar amount of load shedding. The independent runs give different solutions due to the stochastic nature of the GA, 
that can be useful for exploring the solution space and identifying trade-offs between different objectives. The incident managers can use this to their benefit along with domain knowledge to select the best solution. 

In addition, we also investigated optimal load shedding under the case of N-K contingency, and the results showed similar success. Moreover, we analyzed a larger network, CASE9241PEGASE \cite{d1}, and were able to successfully find solutions for it under N-1 contingency.

\section{Conclusions}
In this paper, our approach offers significant benefits, including optimal load shedding recommendations within a quick time frame, flexibility in choosing loads for shedding and the ability of partial load shedding to minimize overall impact. Also, the approach assigns importance to critical loads and enables the incident commander to make informed decisions quickly by taking all relevant factors into account. We have shown the efficacy of our proposed approach using the RTS-GMLC test case. Also, we have shown that our approach gives multiple solutions, and along with external information, the incident managers can take the best decision possible. Finally, along with optimal load shedding, our approach help us to understand the electric grids better, as well as determine crucial lines, critical loads, which can lead us to improve the power system stability, reliability, and power quality.

There are several potential areas for further development and exploration. Our solution can be applied at finer granularity for optimal partial load distribution at bus level. Further development can involve testing on various networks, extensive N-K contingency cases for robustness assessment, and collaborating with utility companies for usability feedback.

\bibliographystyle{apalike}
\bibliography{sample}

\begin{thebibliography}{}

\bibitem[Anderson and Bell, 2012]{f5}
Anderson, B. and Bell, M. (2012).
\newblock Lights out impact of the august 2003 power outage on mortality in new
  york, ny.
\newblock {\em Epidemiology (Cambridge, Mass.)}, 23:189--93.

\bibitem[Barrows et~al., 2020]{f6}
Barrows, C., Bloom, A., Ehlen, A., Ikäheimo, J., Jorgenson, J., Krishnamurthy,
  D., Lau, J., McBennett, B., O’Connell, M., Preston, E., Staid, A., Stephen,
  G., and Watson, J.-P. (2020).
\newblock The ieee reliability test system: A proposed 2019 update.
\newblock {\em IEEE Transactions on Power Systems}, 35(1):119--127.

\bibitem[CBC, 2011]{CBC}
CBC (2011).
\newblock Texas wildfire likely caused by power line sparks.

\bibitem[CEC, 2019]{cec}
CEC (2019).
\newblock Southern california edison company’s 2019 wildfire mitigation plan.

\bibitem[Fernando et~al., 2019]{ir1}
Fernando, W., Rankothge, W., Perera, A., Dissanayake, S., and De~Silva, W.
  (2019).
\newblock Optimization of customer-friendly manual load shedding system.
\newblock In {\em ICAC}, pages 476--480.

\bibitem[Fliscounakis et~al., 2013]{d1}
Fliscounakis, S., Panciatici, P., Capitanescu, F., and Wehenkel, L. (2013).
\newblock Contingency ranking with respect to overloads in very large power
  systems taking into account uncertainty, preventive, and corrective actions.
\newblock {\em IEEE Transactions on Power Systems}, 28(4):4909--4917.

\bibitem[Hassan et~al., 2019]{i2}
Hassan, Y., Rashid, Y., and Tuamiah, F. (2019).
\newblock Demand priority in a power system with wind power contribution load
  shedding scheme based.
\newblock {\em Journal of Engineering}, 25:92--110.

\bibitem[Jabian et~al., 2018]{JABIAN2018486}
Jabian, M.~E., Funaki, R., and Murata, J. (2018).
\newblock Load shedding optimization considering consumer appliance
  prioritization using genetic algorithm for real-time application.
\newblock {\em IFAC-PapersOnLine}, 51(28):486--491.
\newblock 10th IFAC Symposium on CPES 2018.

\bibitem[Jazebi et~al., 2020]{f4}
Jazebi, S., de~León, F., and Nelson, A. (2020).
\newblock Review of wildfire management techniques—part i: Causes,
  prevention, detection, suppression, and data analytics.
\newblock {\em IEEE Transactions on Power Delivery}, 35(1).

\bibitem[Jobanputra et~al., 2017]{jalpa}
Jobanputra, J., Kotwal, C., and Thakkar, J. (2017).
\newblock Linear sensitivity analysis and series compensation for transmission
  congestion management.

\bibitem[Kody et~al., 2022]{o3}
Kody, A., West, A., and Molzahn, D.~K. (2022).
\newblock Sharing the load: Considering fairness in de-energization scheduling
  to mitigate wildfire ignition risk using rolling optimization.
\newblock In {\em 2022 IEEE 61st Conference on Decision and Control (CDC)},
  pages 5705--5712.

\bibitem[Miller et~al., 2017]{MILLER2017267}
Miller, C., Plucinski, M., Sullivan, A., Stephenson, A., Huston, C., Charman,
  K., Prakash, M., and Dunstall, S. (2017).
\newblock Electrically caused wildfires in victoria, australia are
  over-represented when fire danger is elevated.
\newblock {\em Landscape and Urban Planning}, 167:267--274.

\bibitem[PG\&E, 2019a]{pg3}
PG\&E (2019a).
\newblock Pacific gas and electric company amended 2019 wildfire safety plan.

\bibitem[PG\&E, 2019b]{pg1}
PG\&E (2019b).
\newblock Press release: Pg\&e files for reorganization under chapter 11.

\bibitem[PG\&E, 2020]{pg2}
PG\&E (2020).
\newblock Press release: Pg\&e reaches plea agreement on state charges related
  to 2018 camp fire.

\bibitem[Rajappa, 2012]{o1}
Rajappa, G.~P. (2012).
\newblock Solving combinatorial optimization problems using genetic algorithms
  and ant colony.

\bibitem[Rao et~al., 2013]{ir2}
Rao, K.~U., Bhat, S.~H., Ganeshprasad, G.~G., Jayaprakash, G., and Pillappa,
  S.~N. (2013).
\newblock A novel grading scheme for loads to optimize load shedding using
  genetic algorithm in a smart grid environment.
\newblock In {\em 2013 IEEE Innovative Smart Grid Technologies-Asia (ISGT
  Asia)}, pages 1--6.

\bibitem[Rhodes et~al., 2021]{o4}
Rhodes, N., Ntaimo, L., and Roald, L. (2021).
\newblock Balancing wildfire risk and power outages through optimized power
  shut-offs.
\newblock {\em IEEE Transactions on Power Systems}, 36(4):3118--3128.

\bibitem[T. et~al., 2019]{r1}
T., J., S., Y., D., M., F., C., and R., T. (2019).
\newblock 2019 wildfire risk report.

\bibitem[Thurner et~al., 2018]{pp}
Thurner, L., Scheidler, A., Schafer, F., Menke, J.~H., Dollichon, J., Meier,
  F., Meinecke, S., and Braun, M. (2018).
\newblock pandapower - an open source python tool for convenient modeling,
  analysis and optimization of electric power systems.
\newblock {\em IEEE Transactions on Power Systems}.

\bibitem[Wosley and Nemhauser, 2014]{b1}
Wosley, L.~A. and Nemhauser, G.~L. (2014).
\newblock {\em Integer and Combinatorial Optimization}.
\newblock Wiley-Intersience.

\bibitem[Zanocco et~al., 2021]{o2}
Zanocco, C., Flora, J., Rajagopal, R., and Boudet, H. (2021).
\newblock When the lights go out: Californians' experience with
  wildfire-related public safety power shutoffs increases intention to adopt
  solar and storage.
\newblock {\em Energy Research \& Social Science}, 79:102183.

\end{thebibliography}

\end{document}